\documentclass[useAMS,usenatbib]{mn2e}

\usepackage{times}
\usepackage{psfig}
\usepackage{graphicx}
\usepackage{natbib}

\def\aap{A\&A}

\def\apj{ApJ}
\def\apjl{ApJL}
\def\aj{Astronom. Journal}

\def\jcap{JCAP}

\def\mnras{MNRAS}
\def\nat{Nature}
\def\na{New Astronomy}
\def\nar{New Astronomy Reviews}

\def\prd{Phys. Rev. D}

\def\ssr{Space~Sci.Rev.}

\def\spose#1{\hbox to 0pt{#1\hss}}
\newcommand\lsim{\mathrel{\spose{\lower 3pt\hbox{$\mathchar"218$}}
         \raise 2.0pt\hbox{$\mathchar"13C$}}}
\newcommand\gsim{\mathrel{\spose{\lower 3pt\hbox{$\mathchar"218$}}
         \raise 2.0pt\hbox{$\mathchar"13E$}}}

\title[Identifying extreme BL Lacs ]{An emerging population of BL Lacs with extreme properties: towards a class of EBL and cosmic magnetic field probes?}

\author[Bonnoli et al.]{G. Bonnoli $^{1}$\thanks{e--mail: giacomo.bonnoli@brera.inaf.it}, F. Tavecchio$^{1}$, G. Ghisellini$^{1}$, T. Sbarrato$^{1,2}$\\
$^{1}$INAF -- Osservatorio Astronomico di Brera, via E. Bianchi 46, I--23807 Merate, Italy\\
$^{2}$Dipartimento di Fisica ``G. Occhialini'', Universit\`a di Milano
  Bicocca, Piazza della Scienza 3, I-20126 Milano, Italy
}
\begin{document}

\date{Accepted .... Received ...; in original form ...}

\pagerange{\pageref{firstpage}--\pageref{lastpage}} \pubyear{....}

\maketitle

\label{firstpage}

\begin{abstract}
High energy observations of extreme BL Lac objects, such as 1ES 0229+200 or 1ES
0347--121, recently focused interest both for blazar and jet physics and for
the implication on the extragalactic background light and intergalactic
magnetic field estimate. However, the number of these extreme highly peaked BL
Lac objects (EHBL) is still rather small. Aiming at increase their number, we
selected a group of EHBL candidates starting from the BL Lac sample of
\citet{plotkin11}, considering those undetected (or only barely detected) by
the Large Area Telescope onboard {\it Fermi} and characterized
by a high X-ray vs. radio flux ratio. We assembled the multi-wavelength spectral energy distribution of the resulting 9 sources, profiting of publicly available archival observations performed by {\it Swift},  {\it Galex}, and {\it Fermi} satellites, confirming their nature. Through a simple one-zone synchrotron self-Compton model we estimate the expected VHE flux, finding that in the majority of cases it is within the reach of present generation of Cherenkov arrays or of the forthcoming Cherenkov Telescope Array (CTA).  
\end{abstract}

\begin{keywords}
galaxies: active -- galaxies: jets -- radiation mechanisms: non--thermal -- gamma-rays: galaxies.
\end{keywords}

\section{Introduction}

Intense emission of $\gamma$ rays is a distinctive feature of blazars, active galactic nuclei 
(AGN) dominated by the boosted non--thermal continuum from a relativistic jet pointed toward the observer. 
The 2LAC catalogue \citep{ackermann11}, listing the AGN detected with high significance by the 
Large Area Telescope onboard {\it Fermi} during its first two years of operations, 
contains 886 sources, of which 862 are blazars (395 sources classified as BL Lacs, 310 Flat Spectrum Radio 
Quasars (FSRQ), and 157 sources of ``unknown type").
Blazars also dominate the extragalactic sky at very high energy (VHE, $E>100$
GeV)  with BL Lac being the dominant population (58 over a total of 67
extragalactic sources
discovered until  November 2014 according to
TeVCat\footnote{http://tevcat.uchicago.edu}). Within BL Lacs, the large
majority (51 out of 58) of
the VHE emitters belongs to the High-peaked BL Lac (HBL) subclass.

The spectral energy distribution (SED) of blazars displays two characteristics 
broad humps, whose peak frequency appears to anticorrelate with the emitted power 
(\citealt{fossati98}, but see \citealt{giommi05}). While the low energy (from radio up to optical) emission is 
clearly associated to synchrotron radiation, the nature of the mechanisms 
responsible for the high-energy continuum is still debated. 
The majority of the studies adopts the so called leptonic scenario, in which the high 
energy radiation is interpreted as the product of the inverse Compton (IC) scattering 
of the relativistic electrons and soft photons (either produced internally, i.e. 
the synchrotron photon themselves, or externally to the emitting region). 
Hadronic models \citep{bottcher13} instead assume that the $\gamma$--ray emission is either the 
byproduct of reactions initiated by ultrarelativistic hadrons and mediated by mesons 
\citep[e.g.][]{mucke03,atoyan03} or direct synchrotron emission 
from protons \citep[e.g.][]{aharonian00}.

\begin{figure*}
\vspace*{-3.5cm}
\includegraphics[width=0.92\textwidth,]{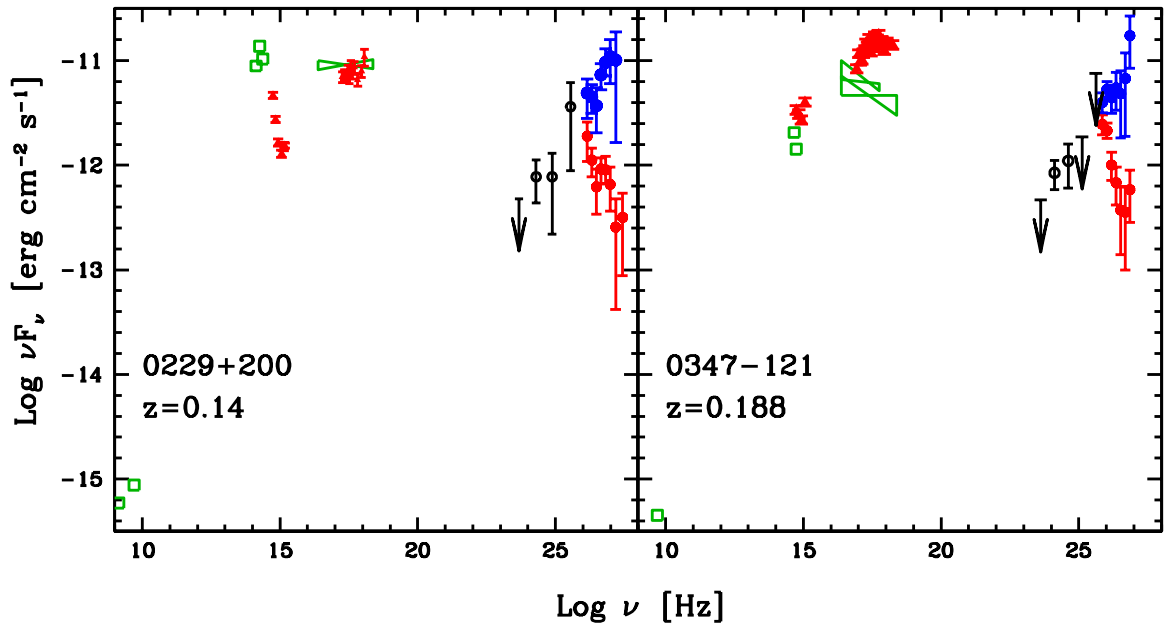}
\vspace*{-3.5cm}
\caption{Spectral energy distribution of 1ES 0229+200 (left) and 1ES 0347-121
   (right), two of the most representative EHBL detected at TeV energies
  (see \citealt{tavecchio11} for references). Blue symbols show the TeV
  spectrum corrected for the absorption by EBL using the model of
  \citet{dominguez11}. The black points for 1ES 0229+200 report the
  \emph{Fermi}/LAT spectrum obtained by \citet{vovk12}, while those for
    1ES 0347-121 come from \citet{tanaka14}.}
\label{known}
\end{figure*}

For the great majority of known BL Lac objects the high-energy component peaks
in the 1-100 GeV band. However, there is a small group of sources detected at
TeV energies by Cherenkov arrays for which the maximum is located above 1 TeV
\citep[e.g.][]{tavecchio11}. The extreme hardness of the spectrum makes these
sources very faint and thus often undetected in the  Large Area Telescope (LAT) band. These features
can be exploited to effectively  constrain the intergalactic magnetic field
\citep[IGMF, e.g.][]{neronov10,tavecchio10,tavecchio11,dolag09,dermer11,vovk12,oikonomou14}. Other properties shared by these
sources are the extremely large ratio between the X-ray and the radio flux and
the hardness of the X-ray continuum ($\Gamma \sim 2$), locating the
synchrotron peak in the medium-hard X-ray band. These characteristics lead to
collect them under the term ``extreme" HBL \citep[EHBL,][]{costamante01}. In
the IR-optical regime the emission is dominated by the host galaxy and the
non-thermal jet continuum starts to be important only in the UV band. The
modeling of their SED  within the Synchrotron Self-Compton (SSC) framework reveals rather unusually low magnetic fields ($B<0.01$
G) and large electron energies \citep{tavecchio10,tavecchio11}. The emitted hard
TeV spectrum (once corrected for interaction with the extragalactic background
light, EBL) can be reproduced assuming that the electron energy distribution
is truncated below a minimum Lorentz factor around $\gamma _{\rm min} \sim
10^5$ \citep{katarzynski05,tavecchio09,tavecchio11,kaufmann11}. This
interpretation also accounts for the peculiar UV/X-ray
spectrum. Alternatively, the SED could be the result of a Maxwellian electron
distribution \citep{lefa11}, internal absorption
\citep{aharonian08,zacharopoulou11} or  inverse Compton scattering  between
electrons and  photons of the cosmic microwave background in the large-scale
($\sim$kpc) jet \citep{boettcher08}.  Alternatively lepto-hadronic models can be also
invoked, either through proton-synchrotron emission, or through secondary
cascades inside the emission region initiated by ultra-relativistic hadrons
\citep[see e.g.][]{cerruti15}. A last suggestive possibility is that
high-energy photons are produced by ultra-high energy protons along the line
of sight injected by the blazars into the intergalactic space \citep{essey11,murase12a,zheng13,tavecchio14}.

From the brief description given above is clear that EHBL are rather
interesting objects, both for the study of jets phenomenology or even Ultra
High Energy Cosmic Ray (UHECR) astrophysics and for the use
of probes of EBL and IGMF. However, their use is somewhat hampered by the
small number of EHBL detected at TeV energies. This is the main fact driving
the present work, in which we intend to characterize a group of EHBL
detectable at TeV energies by present instruments or by the upcoming Cherenkov
Telescope Array \citep[CTA,][]{acharya13}. Sparse groups of EHBL have been identified in previous
work \citep[e.g.][]{giommi05,nieppola06}. Here we intend to follow a focused
and well defined selection procedure, based on the compilation of the
SDSS/FIRST BL Lac of \citet{plotkin11}, and the criterion of an extreme
radio-to-X-ray flux ratio (\S 2). We show their SED using recent {\it Swift}
data (\S 3). We also profit from archival  Galaxy Evolution Explorer
  (GALEX) satellite data and we analyze the whole \emph{Fermi}/LAT photon archive in
order to constrain their behavior in gamma rays by computing at least upper
limits if no detection is possible, which comes out to be the most common
result, in agreement with our expectations. Based on a simple synchrotron
self-Compton homogeneous model we reproduce the resulting SED (\S 4), giving
also an estimate of their flux and detectability for Imaging Atmospheric
Cherenkov Telescopes (IACT) of the present and forthcoming generation. 

Throughout the paper, we assume the following cosmological parameters:
$H_0=70$ km s$^{-1}$ Mpc$^{-1}$, $\Omega_{\rm M}=0.3$,
$\Omega_{\Lambda}=0.7$.  We use the notation $Q=Q_X \, 10^X $ in cgs units.

\section{Selection of TeV candidates EHBL}

The SED of  two  representative EHBL detected in the TeV band, 1ES
0229+200 and 1ES 0347-121, are reported in Fig. \ref{known}\footnote{Note that
  due to a bug in the calculation of the effective area, the X-ray spectrum
  reported in \citet{tavecchio09} was too high. In Fig. \ref{known}  the
  correct spectrum is shown.}. To select EHBL candidates we are guided by two
evident peculiarities of these SED, namely the large X-ray/radio flux ratio
and  a non-thermal optical continuum lower than the thermal contribution from the galaxy.  These two features suggest to select EHBL among the BL Lacs with high X-ray/radio flux ratio whose optical spectrum is dominated by the galaxy. 

\begin{figure*}
\includegraphics[width=0.8\textwidth,]{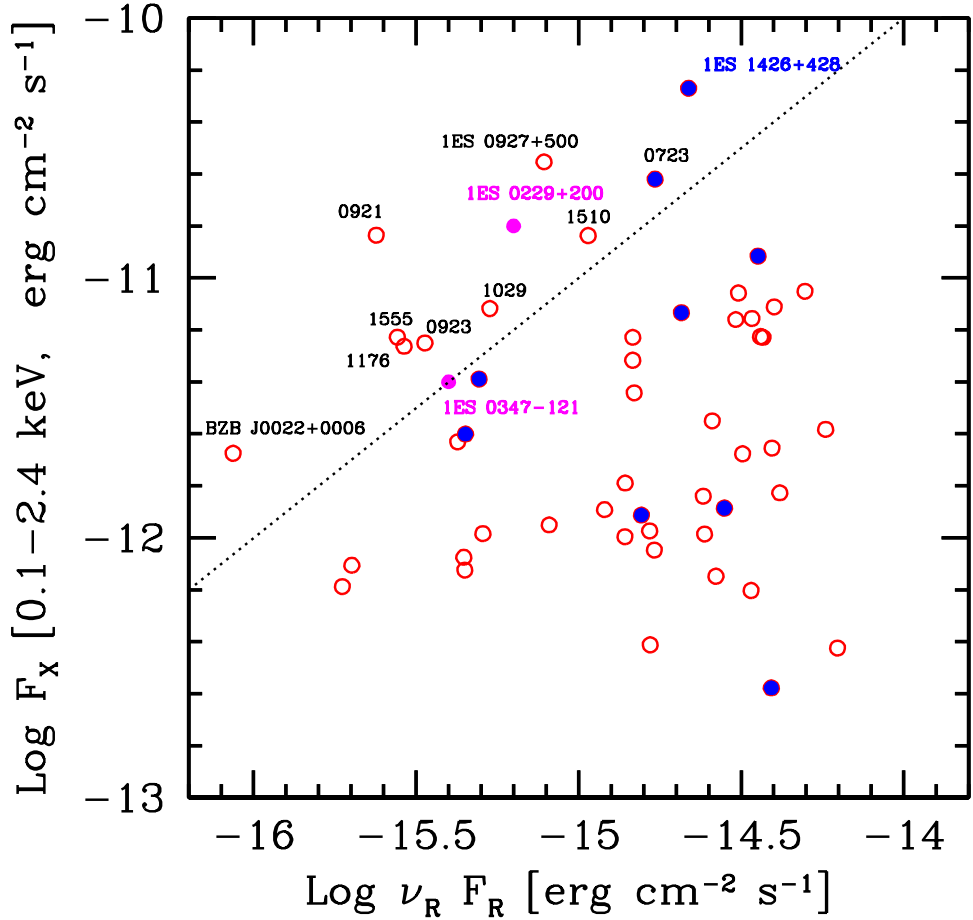}
\vspace*{-0.3cm}
\caption{F$_X$ vs. F$_R$ diagnostic plot for the sample of BL Lacs described in the text. Red open circles represent the full sample; the blue
  filled circles are the LAT detected ones. The 12 sources above the  black dotted diagonal line
  have a high ($F_X/F_R > 10^4$) ratio of X--ray vs. radio  flux. The magenta filled circles are the
  two archetypal TeV detected, but GeV faint extreme BL Lacs, 1ES 0229+200 \citep{vovk12}
  and 1ES 0347-121 \citep{tanaka14}.}
\vspace*{-0.3cm}
\label{frfx}
\end{figure*}

We emphasize that this criterion should not be confused with the one
  adopted in \citet{costamante02} who aimed to maximize the TeV flux (with a selection based on the evidence of large X-ray and radio fluxes), while
  here we tend, in a wide sense, to minimize the TeV spectral index. The
  criterion is rather model independent, as it arises by similarity with the
  archetypal EHBL 1ES 0229+200 and 1ES 0347-121 mentioned above. 
In the scope of a pure one-zone SSC
  model \citep[e.g.,][]{tavecchio98}, this can be interpreted as the outcome
  of  an electron distribution  characterized by high minimum ($\gamma_{\rm min}$) and break ($\gamma_{\rm b}$)  Lorentz factors; this translates into large values of the  peak  frequencies for the synchrotron ($\nu_{s}$) and Inverse Compton
  ($\nu_{IC}$) components of the radiation spectrum, and dim (due to
  scarcity of seed photons) but hard TeV spectra. \citet{costamante02} instead
  requested both high energy electrons and seed photons to be abundant in the
  emission region, in
  order to maximize the bolometric VHE luminosity. This interpretation suggests that the regions of
  the F$_X/$F$_r$ parameter space plotted in Fig. \ref{frfx} selected by
  the two criteria should actually host BL Lac populations of different flavour, with the TeV-brightest clustered towards the upper-right
corner while the TeV-hardest clustered at the upper-left corner. While the
\citet{costamante02} selection has been robustly proved by the detection at
TeV energies of most of their candidate sources, the one we adopt here
will need extensive observation of the selected sources in order to be proven
or rejected.

 Guided by this criterion, we started our selection from the list of 71 BL Lac
presented by \citet{plotkin11} resulting from the correlation of SDSS and
FIRST surveys and by optical spectrum dominated by the host galaxy emission;
to enforce this, we limit the study within $z = 0.4$. To apply our criteria we further select
those with measured X-ray flux (by {\it ROSAT}),  from \citet{plotkin10},
obtaining a total of 50 BL Lacs. Note that the cut in redshift, originally
dictated by the requirement of the galaxy dominance in the optical band, also ensures that the relatively small suppression of the EBL for $\gamma$ rays up to a few TeV does not prevent detection by current
Cherenkov arrays.

The X-ray flux in the 0.1-2.4 keV band and the radio flux at 1.4 GHz
(expressed as $F_{\rm R}\equiv\nu$F$_{\nu}$  for homogeneity with the
integrated X-ray flux), of the sources resulting from the selection have been
taken from \citet{plotkin10} and are reported in Fig.\ref{frfx}. There, the blue points show the sources already
detected by LAT (2FGL catalogue). For comparison, the magenta points report the
fluxes of 1ES 0229+200 and 1ES 0347-121 (although not belonging to our
selected sources). The oblique dashed line corresponds to a fixed ratio $F_{X}/F_{\rm
  R}=10^4$. This particular, rounded but somewhat arbitrary value is
suggested by the position of the two known EHBL in the plane.  The sources above the line are
expected to be good EHBL candidates,  without any claim for completeness, as
  we are not aiming at it at this level. For instance, just outside of the
  region and next to  1ES 0347-121 lies RBS 1049, which would easily fit in
  the same scheme. In this region there are two sources
already detected by LAT, RBS 0723 and 1ES 1426+428. The latter is already
  a well established TeV source \citep{horan02}. Although its SED is coherent with our
picture, it can be considered a transition object towards the ``bright
X-ray''--''bright radio" corner of the plot, populated by the bright blazars
of \citet{costamante02} therefore will be omitted in the following
discussion. It's worth noticing that 1ES1426+428 is not a standalone case, as another hard
source not belonging to our sample but to the one of \citet{giommi05} that is
built along a similar line has been discovered at VHE by HESS \citep[SHBL
  J001355.9-185406 at $z=0.095$,][]{abramowski13}. More recently  RBS 0723
  has been discovered at VHE too, by MAGIC \citep{mirzoyan14}. Other extreme BL Lacs such as
those contemplated in \citet{cerruti15} also show X-ray to radio flux ratios
high enough to fit in the scheme.

In Tab. \ref{Sourcelist} we report
the names of the selected EHBL
together with their coordinates and
redshifts. Note that, as an indirect
result of the selection, requiring a
bright X-ray emission, most of them
belong to the Rosat Bright Survey
\citep[RBS,][]{voges99}. 

\section{Multi--wavelength data}

For the sources fulfilling our criterion on the F$_X/$F$_{\rm R}$ ratio we
built multiwavelength SED exploiting publicly available data. 
We used archival optical--UV data from {\it Swift}/UVOT and soft X--ray data
from  {\it Swift}/XRT for the vast majority of the sources, while for RBS
  0921 and RBS 1176 we requested dedicated {\it Swift} observations. We also profited from the GALEX
  archive\footnote{\tt http://galex.stsci.edu/galexview/}. 
Other data have been also added using the ASI Science Data Center (ASDC) archive\footnote{\tt http://tools.asdc.asi.it/}.
 All the considered sources have been observed at least one time with {\it Swift}, ensuring a good description of the
 crucial UV-X-ray band. No particular attempt was made to seek for strictly
 simultaneous observations, that would need a dedicated observational
 campaign, being these sources generally reputed of secondary importance and
 therefore left aside from any monitoring campaign of some time density. 
 
We also analyzed {\it Fermi}/LAT data. Almost all the sources are undetected
and we could only calculate upper limits.

\begin{table*}
 \centering  
        
       \begin{tabular}{lcccccc}
         \hline
         \hline
           Source Name & R.A.(J2000) &           $\delta$(J2000) & l &b&Redshift & A$_B$ \\
         \hline
	BZB     J0022+0006 &5.5040  &0.1161 &107.18&-61.85&0.306 &0.108\\
	RBS    0723             &131.8039  & 11.5640  &215.46&30.89& 0.198 & 0.093\\
	1ES     0927+500   &142.6566  &49.8404 &168.19&45.71&0.187 &0.073\\
	RBS     0921       &164.0275  &2.8704 &249.28&53.28&0.236 &0.178\\
	RBS     0923       &164.3462  &23.0552 &215.96&63.91&0.378 &0.088\\
	RBS     1029       &176.3963  &-3.6671 &273.11&55.34&0.168 &0.130\\
	RBS     1176       &193.2540  &38.4405 &121.36&78.68&0.371 &0.083\\
	RBS     1510       &233.2969  &18.9081 &29.21&52.05&0.307 &0.210\\
	RBS     1555       &241.3293  &54.3500 &84.35&45.60&0.212 &0.041\\
          \hline
         \hline
         \end{tabular}
         \caption{List of the extreme blazar candidates selected from the
           sample of \citet{plotkin11}. For each source the equatorial
           (J2000) and galactic coordinates are reported (in degrees), the redshift and
         the $A_B$ extinction coefficient from \citep{schlegel98}. As a
         consequence of the selection within SDSS, all the sources lye at high
         galactic latitudes.}
         \label{Sourcelist}
       
         \end{table*}

In the following we describe the analysis performed on the {\it Swift}/XRT
{\it Swift}/UVOT and {\it Fermi}/LAT data.

\subsection{\textit{Swift}/XRT data}

{\it Swift}/XRT  \citep{burrows05} observations were available for all the sources in our
subsample, except for RBS 0923 and RBS 1176, which were targets of dedicated
observations: RBS 0923 was observed in July 2012, while RBS 1176 was observed
in November 2011, then in May and June 2012 . For most of the other sources only one observation was
available, except for RBS 1029  which was observed in November and December
2007,  1ES0927+500, observed  in September 2010 and
March 2011, and RBS 1510 in June,
September, October 2011 and in
January 2012.

{\it Swift}/XRT data were analyzed by using the HEASoft v. 6.13 software package with the CALDB updated on 21 January 2013 and processed with {\tt xrtpipeline v. 0.12.6} with standard parameters. Spectra have been grouped to have at least 20 counts per bin, in order to use the $\chi^2$ test and analyzed with xspec v. 12.8.0 in the 0.3--10 keV energy band. 

 For all sources (with perhaps the exception of RBS 1176, see below) an absorbed power law model provides a good description of the spectrum. In the majority of cases the absorption column
can be fixed to the Galactic value \citep{kalberla05}. 


In Tab. \ref{XRTdata} we report the
results of the fitting procedure. 
 In the two cases of 1ES 0927+500 and
RBS 1029 we separately consider two spectra. The
best fit parameters 
are perfectly consistent. All the other  multiple datasets
could be merged as no hint of
variability was found.

\begin{table*}
 \centering  
        
       \begin{tabular}{llccccc}

\hline
\hline
           Source Name & Obs. ID & Exp. Time& $\Gamma$ & N$_{\rm H}$& $\chi^2/{\rm d.o.f.}$&F$_{0.3-10 \, {\rm keV}}$ \\
& & [s] & & [$10^{20}$ cm$^{-2}$] & & [10$^{-12}$ erg cm$^{-2}$ s$ˆ{-1}$]\\
           \hline
         \hline
	BZB     J0022+0006 &38113001&4700 & $2.40\pm0.25$& 2.76 (Gal.)& 7.19/8& $1.7\pm0.3$\\
	RBS     0723      &37396001      &2000 & $1.78\pm0.12$& 3.17 (Gal.)& 20.8/18& $2.6\pm0.3$ \\
	1ES     0927+500 1  &39154001& 2870 &  $2.0\pm0.1$&1.38 (Gal.) &40.13/37 & $14\pm1$\\ 
	\,\,\,\,\,\,\,\,\,\,\,\,\,\,\,\,\,\,\,\,\,\,\,\,\,\,\,\, 2+3 &39154002-3& 2260 &  $2.2\pm0.1$& 1.38 (Gal.) & 29.4/33 & $16\pm1$\\	
      RBS     0921         &37547001    & 4700 & $1.89\pm0.08$& 3.82 (Gal.) & 44.8/46& $8.4\pm0.4$\\
	RBS     0923           &48001001-4 & 6400& $2.2\pm0.1$& 1.12 (Gal.)& 25.8/26& $3.6\pm0.3$\\
	RBS     1029    1      &36813001&2900& $2.2\pm0.2$& 2.22 (Gal.)& 8.1/6& $2.7\pm0.3$\\
\,\,\,\,\,\,\,\,\,\,\,\,\,\,\,\,\,\,\,\,\,\,\,\,\, 2 &36813002&2900 & $2.0\pm0.1$&  2.22 (Gal.)  & 9.35/10& $3.2\pm0.2$\\
		RBS     1176           &48000001-3&5250 & $2.3\pm0.3$& $13.3\pm6.5^{1}$& 14.1/13& $2.7\pm0.3$\\
	RBS     1510          &91101001-5& 3700 & $2.2\pm0.1$& 3.83 (Gal.)& 35.0/33& $8.3\pm0.6$\\
	RBS     1555          &38303001 & 7100 & $2.0\pm0.1$& 0.886 (Gal.)& 23.4/29& $1.7\pm0.1$ \\    
          \hline
         \hline
         \end{tabular}
         \caption{Results of the {\it Swift}/XRT data analysis for all the archival observations of
           our sample available until the end of August 2012.  The table
           reports:  source name,  exposure time
           (expressed in $s$), best fit value of spectral index of power-law
           model,  column density (in units of $10^{20}$ cm$^{-2}$), $\chi^2$
           and degrees of freedom of the best fit, flux in the 0.3-10 keV band
           (in units of $10^{-12}$ erg cm$^{-2}$ s$^{-1}$). \, $^1$: For this
           source N$_{\rm H, gal}=1.72\times 10^{20}$ cm$^{-2}$.}
\label{XRTdata}
       
         \end{table*}

More complex is the case of RBS 1176. A fit of the summed spectrum with a power law model returns an absorption column $N_H$ largely in excess to the Galactic value (Table 2).  Alternatively, a good fit ($\chi^2/d.o.f.=12.46/13$) can be obtained assuming a broken power law model and the Galactic value of $N_H$, with slopes $\Gamma_1=1.1\pm 0.3$, $\Gamma_2=2.3\pm 0.3$ and break energy $E_{b}=1.3\pm 0.3$ keV.
 The spectra of the single pointings are much less constraining. A single power law model (with photon index $\Gamma\sim 1.7$) with Galactic absorption is barely compatible with the data but a curved spectrum is clearly suggested by the shape of the residuals.

\subsection{\textit{Swift}/UVOT data}

{\it Swift}/UVOT \citep{roming05} is a 30 cm diffraction--limited  optical--UV telescope,
equipped with six different filters, sensitive in the 1700--6500
\AA\  wavelength range, in a 17' $\times$ 17' FoV.
We retrieved from the HEASARC
database the UVOT images in which
our target sources were observed. The maximum angular distance from the
optical axis does not exceed 4' for any source.

For all the sources and the available different observations the  analysis was
performed with  the \texttt{uvotimsum} and \texttt{uvotsource} tasks with a
source region of $5''$. while the background was extracted from a source--free
circular region with radius equal to
$50''$.
The extracted 
 magnitudes were corrected for Galactic
extinction  using the values of \citet{schlegel98} (reported in the
last column of Tab. \ref{Sourcelist} and applying the
formulae by \citet{pei92} for the UV filters, and eventually
were converted into fluxes following \citet{poole05}.

Tab. \ref{UVOTdata} reports the observed Vega magnitudes in the
\textit{Swift}/UVOT $v$, $b$, $u$, $m1$, $m2$, and $w2$ filters, together with
statistical uncertainties. Systematic uncertainties are  never greater than 0.03
mag and therefore dominated by statistical ones in the vast majority of cases.

\begin{table*}
 \centering  
        
       \begin{tabular}{ll|cccccc}
\hline
\hline

  Source Name & Obs. ID & $v$ &$b$ &$u$& $w1$&$m2$&$w2$\\

           \hline
         \hline
	BZB     J0022+0006 &38113001&$19.23\pm0.17$&-&$19.17\pm0.10$&-&-&$19.38\pm0.09$\\
	RBS     0723            &37396001&$17.60\pm0.11$&$18.49\pm0.10$&$17.65\pm0.08$&$17.39\pm0.07$&$17.49\pm0.08$&$17.43\pm0.05$\\
	1ES     0927+500   &39154001-3&-&-&$17.32\pm0.03$&-&-&$17.97\pm0.04$\\
	RBS     0921       &37547001&$18.78\pm0.33$&$19.02\pm0.17$&$18.63\pm0.16$&$18.79\pm0.12$&$18.95\pm0.12$&$18.92\pm0.07$\\
	RBS     0923       &48001001-4&$18.97\pm0.19$&$19.67\pm0.15$&$18.82\pm0.09$&$18.59\pm0.09$&$18.68\pm0.1$&$18.6\pm0.08$\\
	RBS     1029       &36813001-2&-&-&$18.47\pm0.06$&-&-&$18.73\pm0.05$\\
	RBS     1176       &48000001-3&$19.49\pm0.21$&$20.05\pm0.15$&$19.53\pm0.13$&$19.35\pm0.13$&$19.10\pm0.13$&$19.52\pm0.14$\\
	RBS     1510       &91101001-5&-&-&-&$17.54\pm0.04$&$17.51\pm0.05$&$17.65\pm0.04$\\
	RBS     1555       &38303001&$19.37\pm0.11$&-&$19.69\pm0.13$&-&-&$19.99\pm0.11$\\    
          \hline
         \hline
         \end{tabular}
         \caption{{\it Swift}/UVOT \emph{observed} magnitudes for all the archival observations of
           our sample available until the end of August 2012. Statistical
           uncertainties only are reported: systematic error is always within
           0.03 mag and almost generally dominated by statistical ones.}

         \label{UVOTdata}
       
         \end{table*}

\subsection{\textit{Fermi}/LAT data}

Publicly available {\it Fermi}/LAT data were retrieved from the Fermi Science
Support Center (FSSC) and analyzed by means of the LAT Science Tools
v. 9.27.1, together with the Instrument Response Function (IRF) Pass 7 and the
corresponding isotropic and Galactic diffuse background models. Source (class
2) photons in the 0.1--100 GeV energy range, collected until 21 August 2012
and coming from direction within 10
degrees from the nominal position of
the source were selected and
filtered through standard FSSC
quality cuts. Standard analysis
steps were then performed,
eventually adopting the \emph{test
statistic} from \citet{mattox96} to assess the significance
of excess signal in correspondance
with our targets. Besides the target (modeled as a simple power law) and backgrounds, all the 2FGL point
sources in the field were included in the model. 

Most of the sources were undetected
on the whole 0.1--100 GeV energy band. For RBS 1510 (TS=23.0) and
1ES0927+500 (TS=24.5) we had
marginal detections while we could
confirm RBS
0723, already present in both the 1FGL and 2FGL catalogs. Then we computed fluxes in the
1-10 GeV and 10-100 GeV bands,
relaxing the limit for a detection
to TS$>$9 for the sources that were
detected in the full band; however,
none of the others reached this
threshold in either energy bin. In
absence of a measured flux we
computed $2 \sigma $ upper limits, following \citet{rolke05}.
Results are collected
in Tab. \ref{LATdata} and plotted
in Figs. \ref{SEDS1}-\ref{SEDS3}.
In the last two columns we also
report the number of photons with
energy E$>$ 10 GeV detected within a
0.4$^\circ$ radius (roughly corresponding to
68\% containment for E $>$ 10 GeV) from the nominal
position of the source, and the
energy of the most energetic one
E$_{max}$. A rigorous study of the
significance of these photons,
taking into account the different
PSF of LAT for front and
back converted photons,
aiming to check and exclude contamination from
nearby hard sources (though
unlikely) and eventually to assess
the probability of enclosing background
photons within the same
aperture, was beyond the scope of
this work.
\begin{table*}
\centering
\begin{tabular}{lccr|cc|cc}
\hline
\hline
Source Name & F$_{0.1-100{\rm GeV}}$
& $\Gamma$& TS& F$_{1-10{\rm GeV}}$
& F$_{10-100{\rm GeV}}$
& Hard photons&E$_{max}$\\  
&[10$^{-10}$ ph cm$^{-2}$
   s$^{-1}]$&&&[10$^{-10}$ ph
     cm$^{-2}$ s$^{-1}$] &
   [10$^{-10}$ ph cm$^{-2}$
     s$^{-1}]$&[$N_{E > 10 {\rm GeV}}$]&[GeV]\\
\hline     
\hline

BZB     J0022+0006    &-&-&1.88&$<1.7$&$<0.4$&1&31.5\\
RBS     0723             &11.3$\pm2.6$&1.46$\pm0.08$&79&3.9$\pm$1.0&0.69$\pm$0.25&10&77.6\\
1ES     0927+500        &4.2$\pm$3.4&1.38$\pm$0.3&24.5&$<1.4$&0.62$\pm$0.26&9&159.1\\
RBS     0921            &-&-&4&$<2.4$&$<0.44$&2&23.4\\
RBS     0923            &-&-&3&$<1.5$&$<0.36$&1&53.7\\
RBS     1029           &-&-&0.002&$<2.0$&$<0.29$&0&-\\

RBS     1176          &-&-&-0.04&$<2.3$&$<0.31$&1&13.7\\
RBS     1510           &5.8$\pm$3.6&1.48$\pm$0.21&23.0&2.0$\pm$0.8&0.36$\pm$0.20&5&42.9\\
RBS     1555            &-&-&0.44&$<2.0$&$<0.39$&2&10.9\\
   
\hline
\hline

\end{tabular} 
\caption{Results of our analysis of
  the {\it Fermi}/LAT data collected until 21
  August 2012 for the sources of our
  EHBL sample. For each target in the first three columns the integral flux
  above 0.1 GeV, the slope of the simple power law model and the TS are
  reported. Then the value of the flux or the  $2 \sigma $ U.L. in the 1-10 GeV and 10-100
  GeV energy bins are reported. In the last two columns, the number of photons
with E $>$ 10 GeV observed within 0.4$^\circ$ from the nominal position of the
source, and the energy E$_{max}$ the most energetic one.}
\label{LATdata}
\end{table*}

\subsection{GALEX data}

The Galaxy Evolution Explorer \citep{martin05} was a 
NASA Small Explorer, in flight since 28 April 2003, and operational until mid 2013.

It performed an all--sky
survey in the far UV (FUV, $\lambda_{{\rm eff}}\sim 154$ nm) and near UV
(NUV, $\lambda_{{\rm eff}}\sim 232$ nm) band.
We retrieved from the MAST\footnote{Multimission Archive at the Space Telescope
Science Institute, \texttt{http://galex.stsci.edu}}  archival fluxes
observed for the sources of our sample (see Tab. \ref{GalexTab}). No data is
found for RBS 1176, likely due to a gap in the sky coverage of the survey. 

\begin{table}
\centering
\begin{tabular}{lccrrr}
\hline
\hline
Source Name & FUV Flux  & NUV Flux \\
            &   [$\mu$Jy] &  [$\mu$Jy]\\
\hline     
\hline

BZB     J0022+0006    &17.0$\pm$0.2&23.40$\pm$0.25\\
RBS     0723          &104.3$\pm$4.9&177.2$\pm$4.3\\
1ES     0927+500      &59.3$\pm$5.6&68.8$\pm$4.1\\
RBS     0921          &15.1$\pm$1.5&25.4$\pm$1.1\\
RBS     0923          &12.6$\pm$3.7&22.8$\pm$3.5\\
RBS     1029          &17.5$\pm$4.3&32.9$\pm$4.9\\
RBS     1176          &-&-\\
RBS     1510          &24.9$\pm$6.2&48.1$\pm$5.3\\
RBS     1555          &9.62$\pm$0.51&11.3$\pm$0.2\\
   
\hline
\hline

\end{tabular} 
\caption{Archival GALEX fluxes in the FUV and NUV bands for our sample of
  EHBL. No data is found for RBS 1176, likely due to a gap in the sky coverage of the survey.  }
\label{GalexTab}
\end{table}

\section{Spectral Energy Distributions}

\begin{figure*}

\includegraphics[width=0.495\textwidth,]{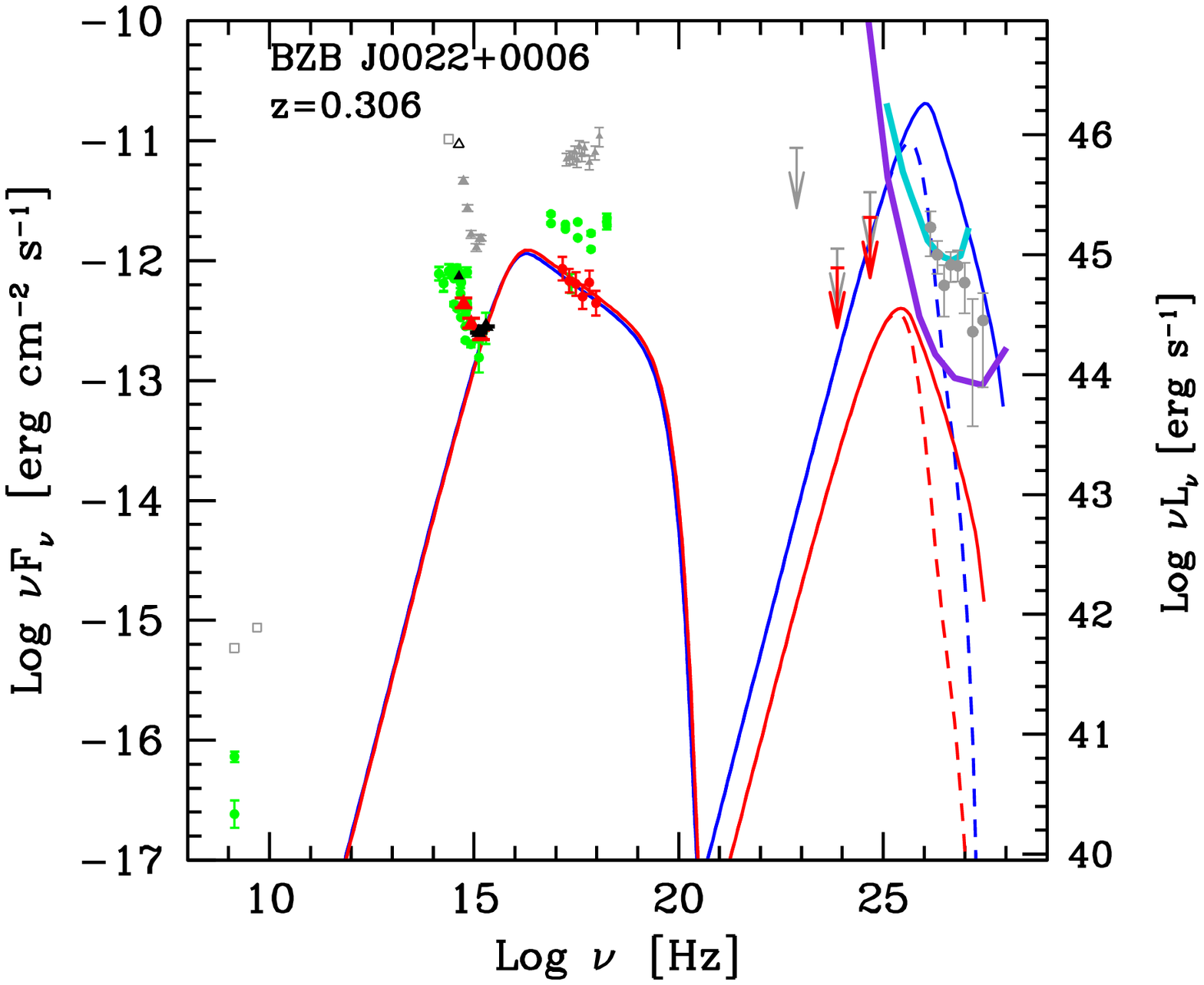}
\includegraphics[width=0.495\textwidth,]{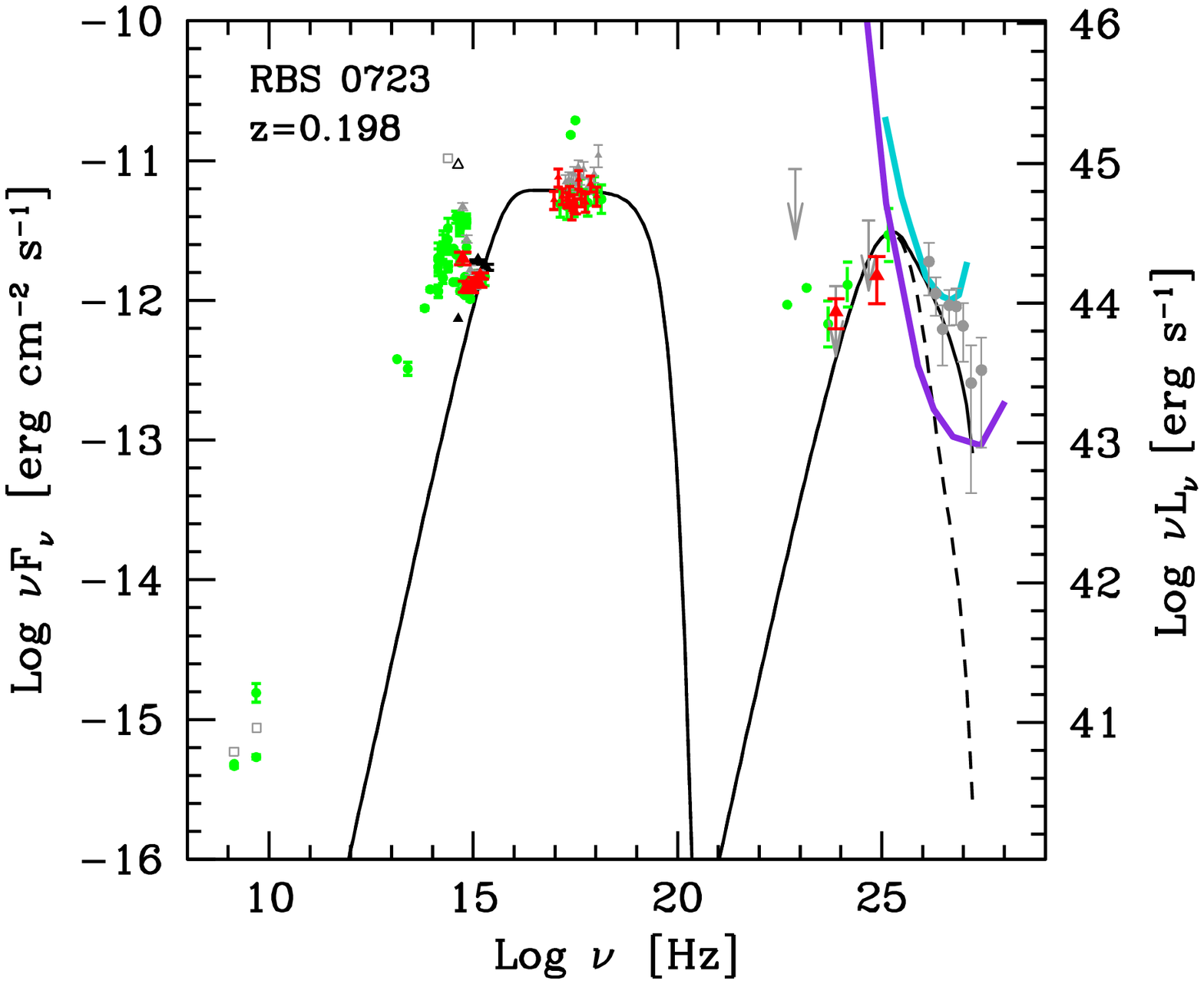}
\includegraphics[width=0.495\textwidth,]{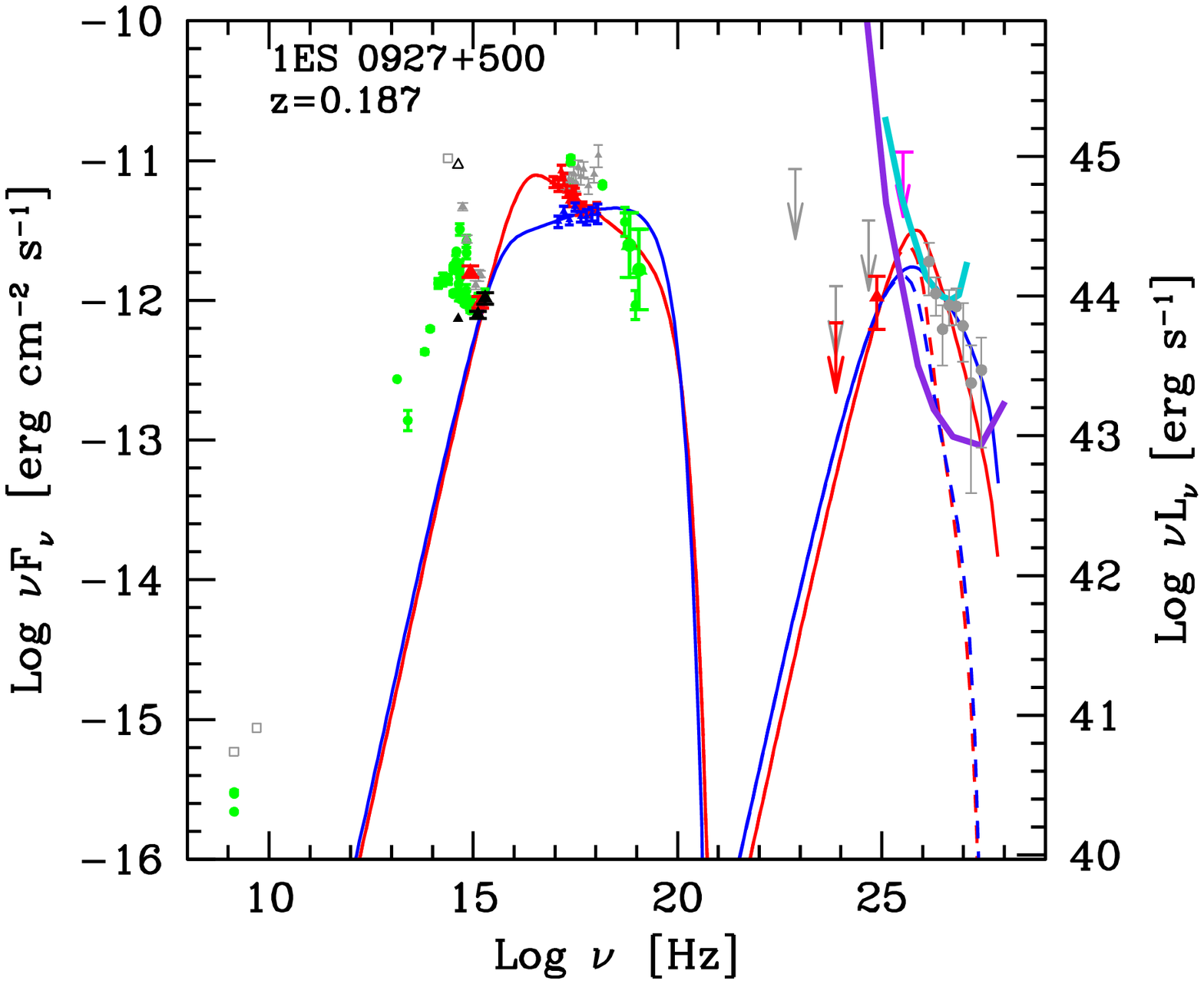}
\includegraphics[width=0.495\textwidth,]{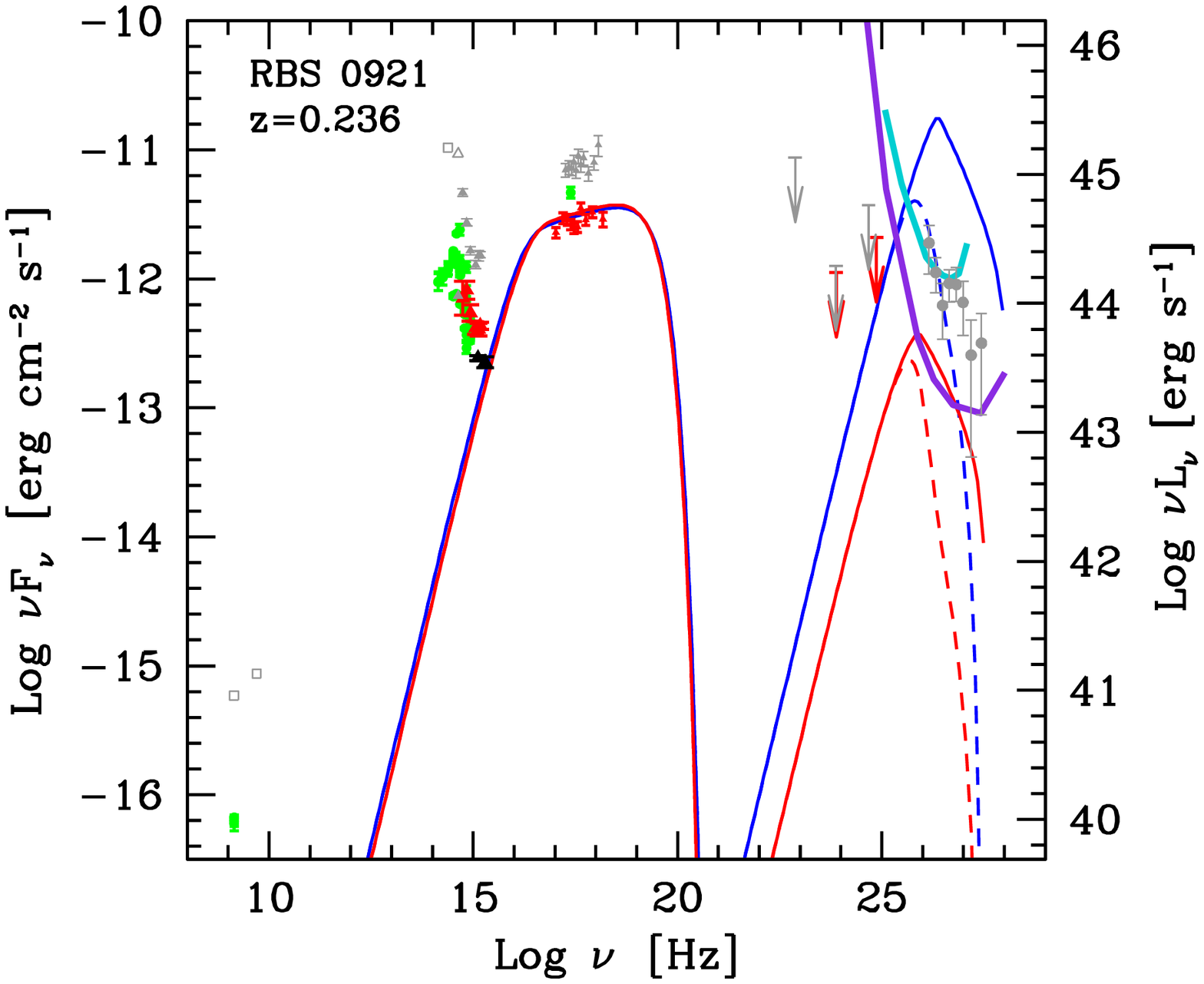}
\caption{Spectral Energy Distribution of the selected extreme BL Lac
  objects. Green symbols reports historical data from ASI/ASDC archive. Red
  symbols show {\it Swift}/UVOT, XRT and {\it Fermi}/LAT data discussed in the
  text. Black symbols display  {\it Galex} data. Background grey symbols show
  the SED of 1ES 0229+200 for comparison. For sources not detected by LAT we
  report two different models of the SED, corresponding to low ($B=0.01$ G,
  blue lines) and high ($B=0.1$ G, red lines) magnetic field. For
  sources with a LAT detection we only report one model, in black (parameters
  are reported in Tab. \ref{param}). For 1ES 0927+500 we report two models, corresponding to the 
  two X-ray slopes (see Tab. 2). Dashed lines show the model after
  absorption with the EBL, calculated according to \citet{dominguez11}. Light
  blue and violet curves report the differential sensitivities ($5\, \sigma$, 50
  hours of exposure, 5 bins per energy decade) of MAGIC and CTA respectively.}
\label{SEDS1}
\end{figure*}

\begin{figure*}
\includegraphics[width=0.495\textwidth,]{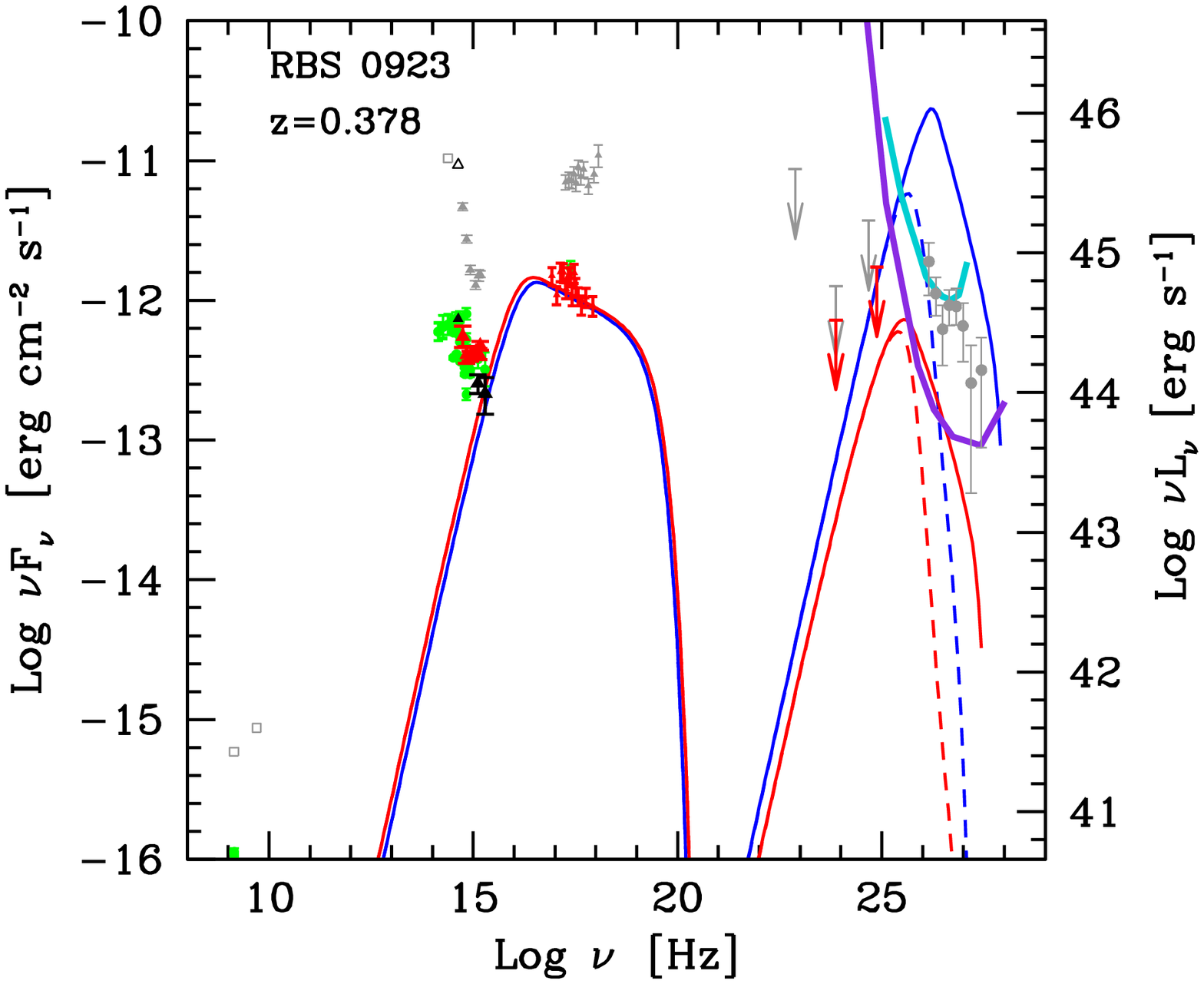}
\includegraphics[width=0.495\textwidth,]{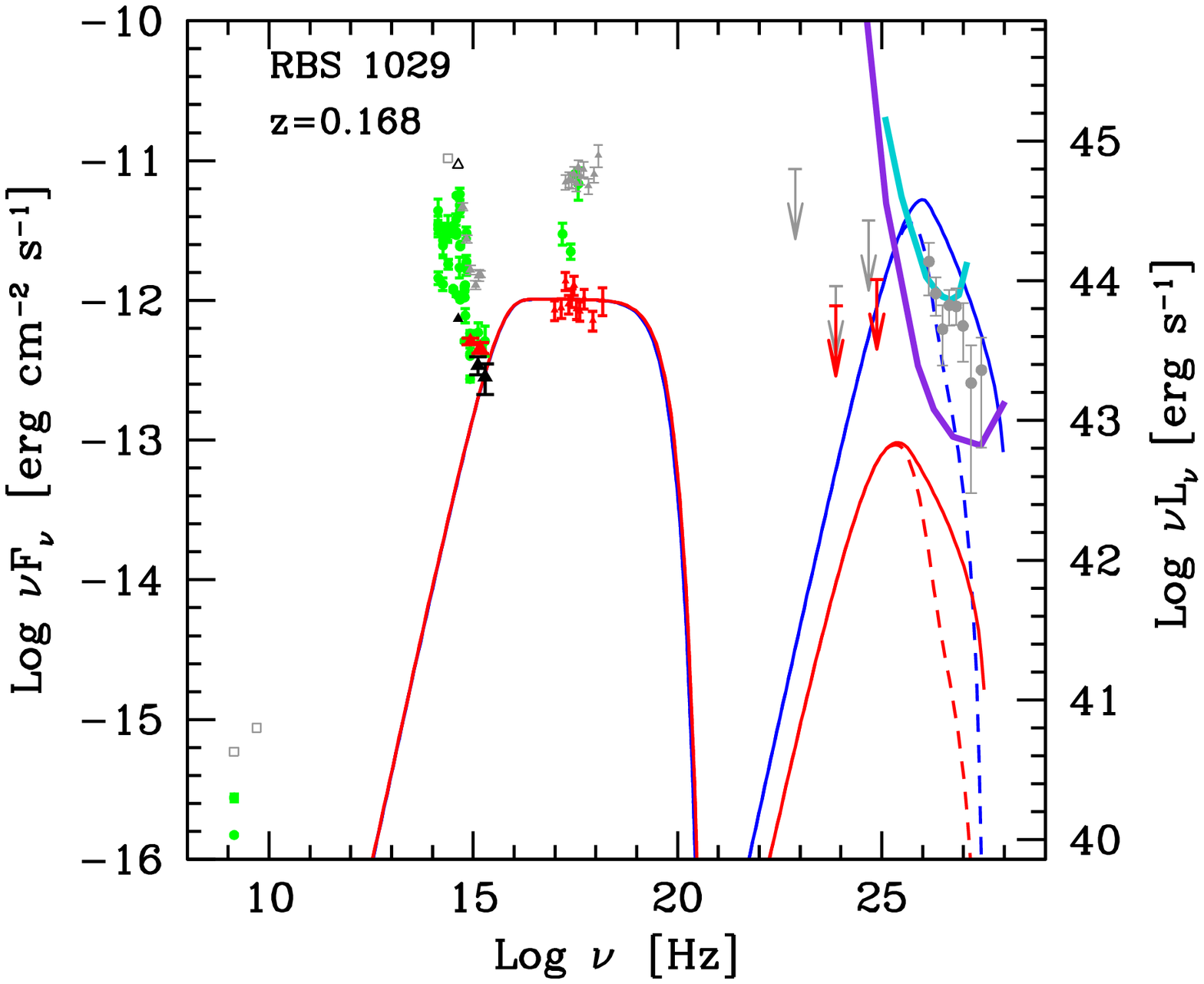}
\includegraphics[width=0.495\textwidth,]{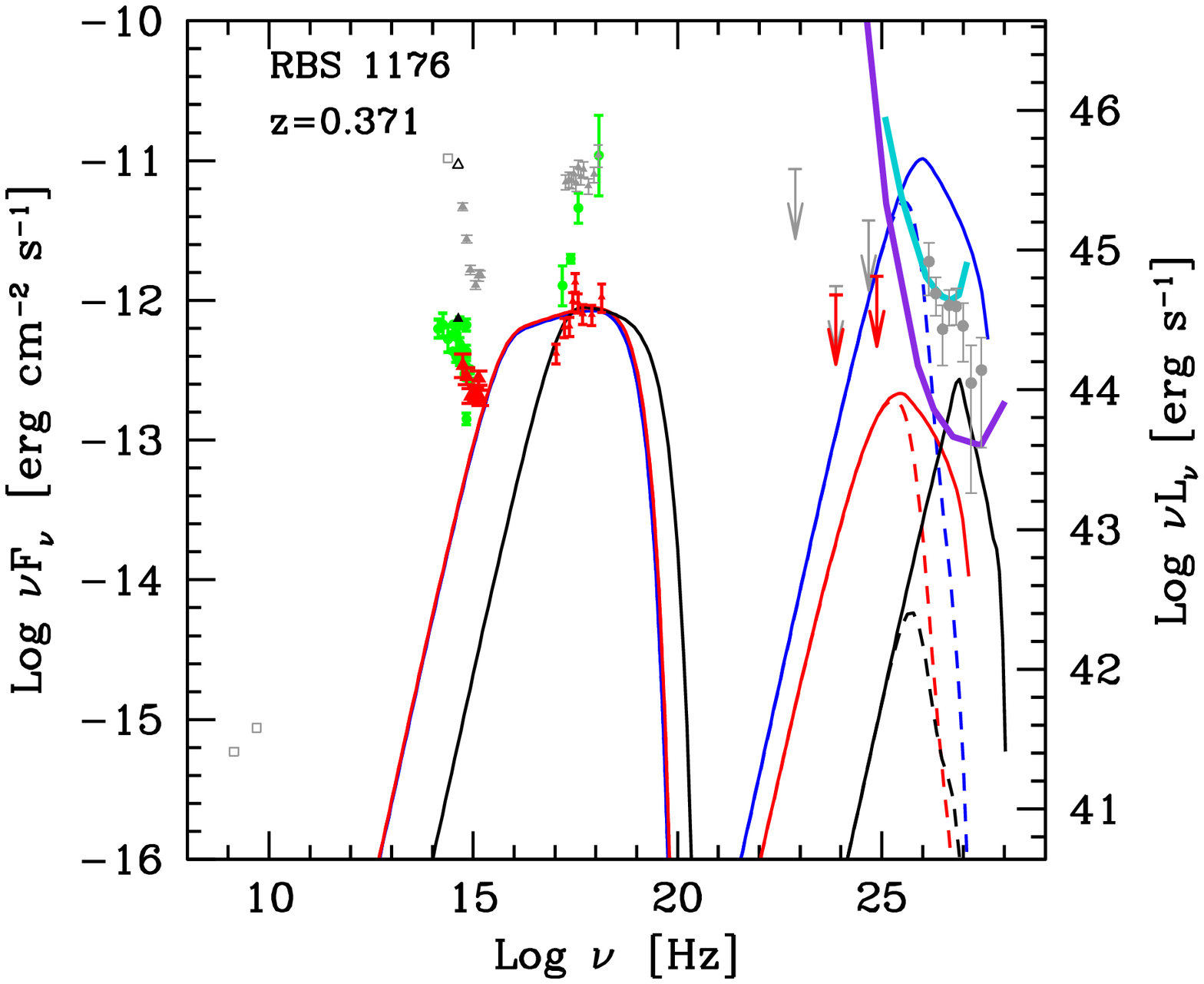}
\includegraphics[width=0.495\textwidth,]{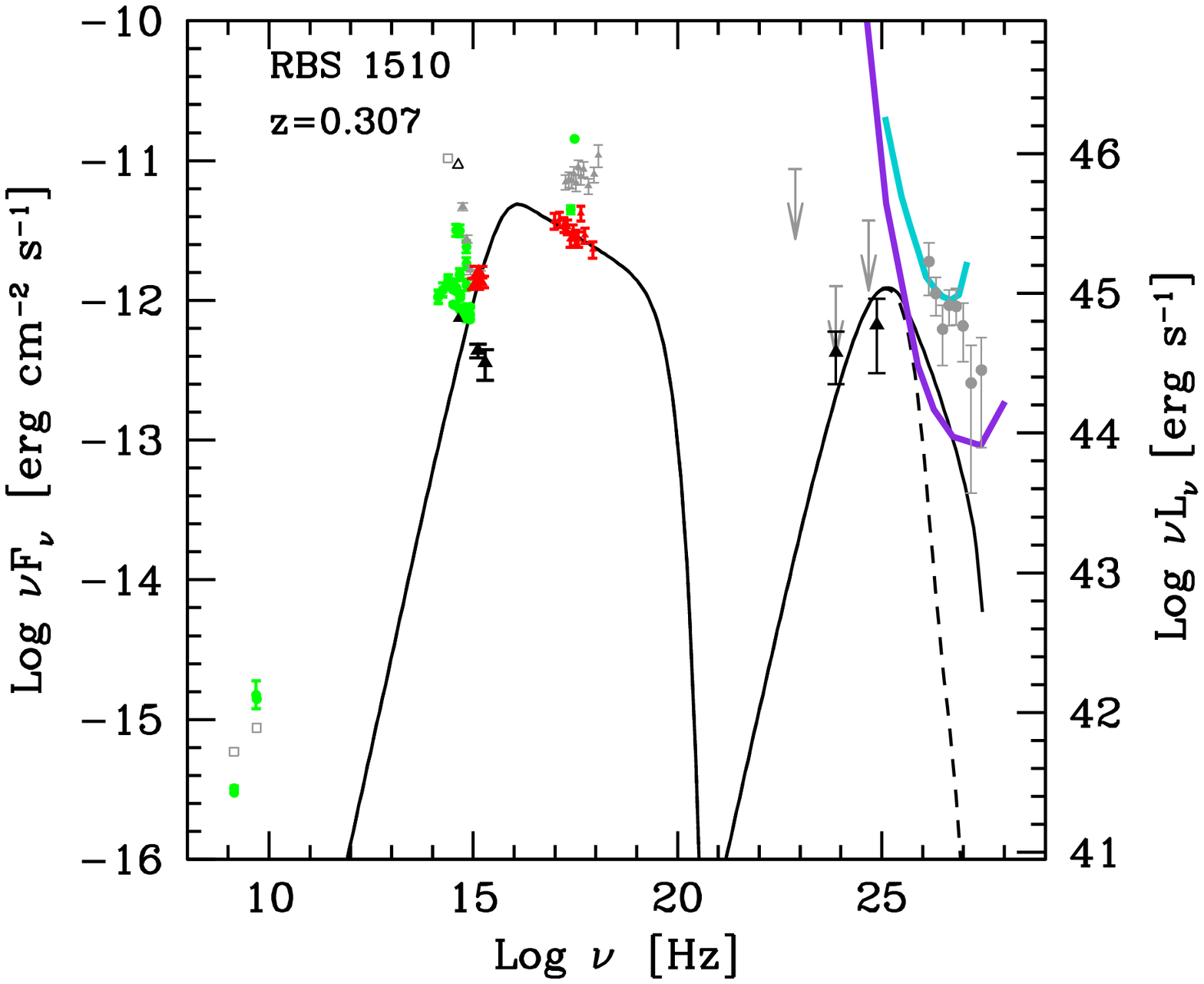}
\caption{As in Fig. \ref{SEDS1}. In the case of RBS 1176, the black line
    refers to the model assuming the highest value of $\gamma_{min}$
    compatible with the X-ray spectrum ($\gamma_{min}=4\times 10^{5}$, see Tab. \ref{param}).}
\label{SEDS2}
\end{figure*}

The SED of the sources are reported in Figs. \ref{SEDS1}-\ref{SEDS3}. In all
cases we use the same color code: green symbols are used for archival data
from  ASDC, red points for {\it Swift}/XRT, UVOT and LAT data and black
triangles for {\it GALEX} data (taken from the database). Note that in almost
all cases the GALEX and UVOT data in the UV filters perfectly agree, in spite
of the uncorrelated observing epochs. The only
exception is RBS 1510 for which GALEX provides fainter fluxes than UVOT. In this case variability is likely an explanation for this difference.

For comparison, we report in background (gray) the data corresponding to 1ES
0229+200 (for simplicity only the observed TeV spectrum is shown). It is clear
that the structure of the synchrotron part of the SED of all the sources
closely resembles that of 1ES 0229+200. In all cases the steep optical
continuum is dominated by the host galaxy emission. A feature shared by all
the sources is the large ratio between the flux in the UV band and soft X-ray band. As remarked in \citet{tavecchio09} \citep[see also ][]{kaufmann11}, in the framework of the one-zone leptonic model, this features can be reproduced if the energy distribution of the emitting electrons is truncated below a relatively large value, $\gamma _{\rm min}=10^{4-5}$. As for 1ES 0229+200, this also causes a very hard SSC component, consistent with the extremely low flux in the LAT band. 

 As discussed in \S 3.1, for RBS 1176 the observed deficit of soft X-ray photons in the XRT spectrum can be interpreted either as due to intrinsic or intervening absorption or as the evidence for an intrinsic curvature of the spectrum. In the latter case, the extremely hard soft X-ray continuum (photon index close to 1) would be readily interpreted in the SSC scheme as the synchrotron low-energy tail of the electrons with Lorentz factor  $\gamma_{\rm min}$. In this case, RBS 1176 would thus be a source characterized by a peculiarly large $\gamma_{\rm min}$, for which the break in the synchrotron continuum -- which in the other sources occurs between the UV and the X-ray bands -- is located around 1 keV. Therefore RBS 1176 could be the first example of {\it ultra extreme HBL}.

 Our selection criterion is rather efficient in selecting EHBL blazars, as
  shown by the similarity of the synchrotron SEDs of the selected sources,
  even if the paucity of the $\gamma$-ray data does not allow (yet) a detailed
  comparison of the high energy SEDs.
 We remark that the selection procedure does not explicitly require that the synchrotron peak is in the X-ray band
(which is the defining feature of EHBL). Nevertheless, almost all the selected
BL Lacs show hard X-ray spectra, often showing photon indices below 2,
locating the synchrotron peak energy above 10 keV; it is worth noticing that
with indices around 2 the position of the synchrotron peak can shift a lot
(from $\sim 10^{16}$ to $\sim 10^{19}$ Hz ), even if the synchrotron
luminosity is almost constant. As shown below, the exact location of the
synchrotron peak does not affect much the predicted TeV flux, since this
depends mostly on the low energy edge of the electron distribution (see the
case of 1ES 0927+500 below).

The high energy component is unconstrained in almost all the sources, with the
exception of RBS 0723, RBS 1510 and 1ES 0927+500, detected by LAT at the
highest energies. Based on the striking similarity of the synchrotron bump, it
is reasonable to assume that the high energy component is also similar to that
of 1ES 0229+200, characterized by an extremely hard spectrum in the LAT {\bf
  ($\Gamma= 1.36 \pm 0.25$ according to \citealt{vovk12}, but see \citealt{cerruti15}
where a softer slope is found)} and TeV bands. This assumption is
supported by the detection of 1ES 0927+500 and RBS 1510 only at high energy
and by the hard photon index of RBS 0723 reported in the 2FGL, $\Gamma
=1.48\pm 0.16$. In the next paragraph we then estimate the expected TeV flux
assuming as SED template and physical parameters those of 1ES 0229+200.

\subsection{SED Modeling and predicted TeV fluxes}

We use the one-zone leptonic model of \citet{tavecchio98}, fully described in \citet{maraschi03}. The emitting region is assumed to be spherical with radius $R$, filled by a tangled magnetic field of intensity $B$. To reduce the number of free parameters we assume that the relativistic electrons follow a simple power law distribution $N(\gamma)=K\gamma ^{-n}$ with $\gamma _{\rm min}<\gamma <\gamma _{\rm max}$. As for the case of 1ES 0229+200 \citep{tavecchio09,kaufmann11} this distribution suitably describes the observed SED. The relativistic beaming of the synchrotron and SSC radiation is described by the relativistic Doppler factor $\delta$. These parameters fully specify the model, that can be uniquely fixed once the quantities specifying the two bumps (peak frequencies and luminosities, spectral slopes) and the variability timescale are known \citep{tavecchio98}. 

In the present case, since in the majority of cases we do not have any
direct measurement of the high energy hump, it is not possible to
uniquely derive the physical parameters of the emitting sources. Moreover the
minimum variability timescale $t_{var}$ for these sources is not known, thus
allowing additional degeneracy in the $K - R$ plane. On the other
hand, relying on the physical parameters inferred  for known EHBL
\citep[e.g.][]{tavecchio11}, it is possible to derive the SSC component and
thus an estimate of the expected TeV flux. 
According to the results of the models in \citet{tavecchio10,tavecchio11} we then fix the radius $R=6\times 10^{15}$ cm and the Doppler factor $\delta=20$. Then we consider two values of the magnetic field bracketing the expected range of the magnetic field intensity, $B=0.01-0.1$ G and for each value we derive the remaining parameters ($n$, $K$, $\gamma _{\rm min}$ and $\gamma _{\rm max}$) reproducing the synchrotron component (described by the X-ray spectrum and the UV data) and the upper limits in the GeV band. 
The only parameter loosely constrained in this procedure is $\gamma _{\rm
  max}$, which, however, has only a minor impact on the derived SSC component
due to Klein-Nishina suppression. For 1ES 0927+500, RBS 0723 and RBS 1510, the
LAT data allow us to constrain also the level of the SSC component and to
determine all the parameters. Therefore in this case we present only one model.
 For RBS 1176, as discussed above, we consider both possibilities for the X-ray spectrum. For the case assuming additional absorption we report two models, as for the other sources. For the case in which the X-ray continuum as an intrinsic curvature we show one model (black lines), assuming $B=0.01$ G. We do not consider the case of a larger magnetic field for which the SSC luminosity (already low) would be much smaller.

As discussed in \citet{katarzynski05} and \citet{tavecchio09}, the peculiar
SED of EHBL can be reproduced in the framework of the standard one-zone
leptonic model assuming that the emitting relativistic electrons follow an
energy distribution truncated below a relatively large energy or,
equivalently, Lorentz factor $\gamma _{\rm min}$. In this case, below the
typical synchrotron frequency of electrons with Lorentz factor $\gamma_{\rm
  min}$, $\nu _{\rm min}\simeq 2.8\times 10^6 B \gamma _{\rm min}^2 \delta$,
the resulting spectrum is described by the characteristics hard power law
$F(\nu)\propto \nu^{1/3}$. The same hard spectrum describes the SSC emission
up to the peak energy  $h \nu _{\rm C}\simeq \gamma _{\rm min} m_{\rm e} c^2
\,\delta$. Since the UV and X-ray data constrain $\nu _{\rm min}$ around
$10^{16}$ Hz, the typical minimum Lorentz factor is $\gamma _{\rm min}\simeq
2\times 10^5 \nu_{\rm min,16}^{1/2} \, B_{-2}^{-1/2} \delta _{1}^{-1/2}$,
implying a SSC peak at $\nu _{\rm C} \simeq 1 \, \nu_{\rm min,16}^{1/2}
B_{-2}^{-1/2} \delta _{1}^{1/2}$ TeV. Note that in this scheme the sources
with very large separation between the X-ray and the UV fluxes, implying a
large value of $\nu _{\rm min}$, are expected to have the maximum of the SSC
component at very high energy and are thus the most promising sources for TeV
detection. Larger values of $\delta$ have been sometimes required in the past
to model the SED of HBL, \citep[e.g.][]{aleksic12}, and would boost even more the
detectability of our sources. 

\subsection{Results}

\begin{figure}
\includegraphics[width=0.495\textwidth,]{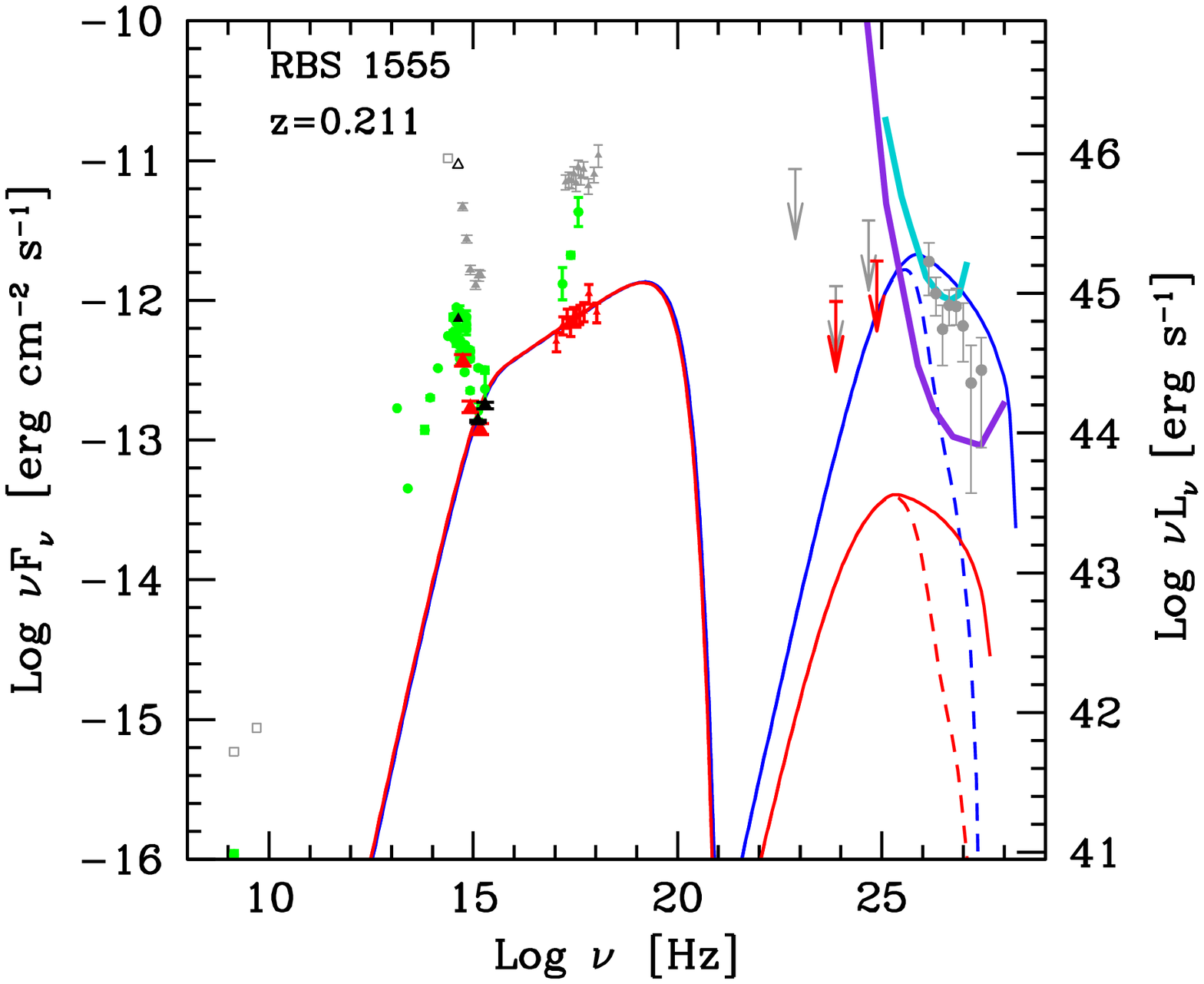}
\caption{As Fig. \ref{SEDS1}.}
\label{SEDS3}
\end{figure}

The resulting theoretical SED are shown by the red ($B=0.1$ G) and blue ($B=0.01$ G) lines in Figs. \ref{SEDS1}-\ref{SEDS3}. Solid lines report the intrinsic emission, dashed lines show the observed emission, corrected for absorption through interaction with the EBL using the model of \citet{dominguez11}. We note here that the radio emission can not be explained by our model, providing fluxes well below the measured level. This is a general feature of single-region models, which are especially intended to model the emission at higher frequencies emitted by compact components. The low frequency radio emission is instead likely produced in extended regions in the jet.

The cases with low $B$ are characterized by a larger SSC flux and {\it viceversa}. This is simply due to the well known fact that the ratio between the SSC and synchrotron luminosities is proportional to the radiation and magnetic energy densities, i.e. $L_{\rm SSC}/L_{\rm syn}=U_{\rm rad}/U_{B}$. For a fixed  $L_{\rm syn}$ also $U_{\rm rad}$ is constant, thus $L_{\rm SSC}\propto B^{-2}$.  

Figs.  \ref{SEDS1}-\ref{SEDS3} also display the sensitivity curves for MAGIC
 \citep[light blue,][]{sitarek13} and CTA \citep[violet,][]{actis11} corresponding to 50 hours of observation and
$5\, \sigma$ significance. It's noteworthy that
these sensitivity curves assume a
0.2 dex energy binning, therefore dimmer flux densities are still within reach
if a more coarse binning is adopted. In the low magnetic field case, the
majority of the sources could be already detected by the present generation of
IACTs and all the sources could be easily detected by CTA
\citep{acharya13}. In the high B-field case, instead, the selected EHBL could
be hard to detect even by CTA. However, we remark  that the prediction of the
SSC flux for the case of high magnetic field, $B=0.1$ G, should be considered
rather pessimistic, since the magnetic field intensity derived for most of the known EHBL tends to lie close to the low value $B=10^{-2} G$ \citep[e.g.][]{tavecchio11}.

For RBS 1176, the case of an intrinsic X-ray break results in a large
  peak frequency (which we recall is directly related to $\gamma_{\rm min}$,
  that in this case is larger than for the other sources) and a quite low SSC
  luminosity (related to the small energy density of the target optical-IR photons). Therefore, if the observed lack of soft photons is really connected  to the intrinsic spectrum RBS 1176 (making this source the first ultra-extreme HBL) we do not expect that it can easily detected by the CTA.

\begin{table*}
 \centering         
       \begin{tabular}{lccccc}
         \hline
         \hline
           Source Name & $B$ & $K$ & $\gamma _{\rm min}$  & $\gamma_{\rm max}$
           & $n$ \\
&[G]&&&&\\
         \hline
        BZB     J0022+0006 & 0.1 & $1.7\times 10^{11}$& $3\times 10^4$&   $2\times10^6$&3.5   \\
                                       & 0.01 & $2.75\times10^{13}$& $9\times 10^4$&   $6\times10^6$&3.5   \\
     RBS     0723              & 0.15 & $6\times 10^{8}$& $2.1\times 10^4$&   $1.5\times10^6$&3.0   \\
        1ES     0927+500$^1$   1    & 0.05 & $1.7\times 10^{10}$& $4.1\times 10^4$&   $3\times10^6$&3.3   \\
        $\;\;\;\;\;\;\;\;\;\;\;\;\;\;\;\;\;\;\;\;$     2+3       & 0.035 & $1.3\times 10^{7}$& $2.7\times 10^4$&   $3\times10^6$&2.7   \\                    
        RBS     0921          & 0.1 & $8\times 10^{7}$& $4.7\times 10^4$&   $1.8\times10^6$&2.8   \\
                                       & 0.01 & $6\times 10^{9}$& $1.3\times 10^5$&   $6\times10^6$&2.8 \\
        RBS     0923         & 0.1 & $1.4\times 10^{8}$& $2.2\times 10^4$&   $2\times10^6$&3.3  \\
                                       & 0.01 & $5.1\times 10^{12}$& $1.2\times 10^5$&   $5.2\times10^6$&3.3   \\
        RBS     1029        & 0.1 & $1.4\times 10^{8}$& $2.2\times 10^4$&   $2\times10^6$&3.0   \\
                                       & 0.01 & $1.4\times10^{10}$& $7\times 10^4$&   $6\times10^6$&3.0   \\
        RBS     1176$^{^2}$       & 0.1 & $6\times 10^{7}$& $2.3\times 10^4$&   $10^6$&2.8   \\
                                       & 0.01 & $4.6\times 10^{9}$& $7.3\times 10^4$&   $3\times10^6$&2.8   \\
                                       & 0.01 & $3\times 10^{11}$& $4\times 10^5$&   $3\times10^6$&3.1   \\
        RBS     1510$^3$        & 0.12 & $6.2\times10^{9}$& $2\times 10^4$&   $2\times10^6$&3.35   \\
        RBS     1555        & 0.1 & $1.2\times 10^{6}$& $1.3\times 10^4$&   $3\times10^6$&2.6   \\
                                           & 0.01 & $7.5\times 10^{7}$& $4.3\times 10^4$&   $10^7$&2.6   \\
          \hline
         \hline
         \end{tabular}
         \caption{Physical parameters describing the jet emission in our
           sample whithin the framework of the SSC model from \citet{tavecchio98}, with the
           simplifying assumption that the energy spectrum of the emitting
           electrons follows a simple (instead of broken) power law distribution. A common Doppler factor
           $\delta=20$ is assumed for all the sources. For each
           target source the intensity of the magnetic field within the emission
           region $B$, the electron spatial density $K$, the minimum and
           maximum electron energy $\gamma_{min}$ and $\gamma_{max}$, the
           slope of the power law distribution $n$ are reported. $^1$: $R=6.5\times 10^{15}$ cm, $\delta=30$;  $^2$ the first two lines refer to the model assuming a power law X-ray spectrum. The third line is for a broken-power law X-ray spectrum (black line in Fig. 4); $^3$:$R=1.4\times 10^{16}$ cm.}
         \label{param}
       
         \end{table*}

\section{Discussion}

Extreme HBLs are particular interesting for several reasons, ranging from the
study of jet physics to the estimate of EBL and IGMF. In this work we have
assembled a well defined group of EHBL and we have characterized their SED
using historical, {\it Swift}, {\it Fermi}/LAT and (when available) {\it
  GALEX} data. All the SEDs closely resemble in shape that of the
``prototypical" EHBL 1ES 0229+200, demonstrating the efficiency of our
selection in identify these kind of sources. We applied a simple one-zone
synchrotron SSC model, trying to predict the flux at very high energies,
showing that most of the sources could be detectable by the upcoming
CTA. Given the shape of the SED, the most effective instruments of the CTA to
detect EHBL would be the  array of Medium Sized Telescopes (MST),
  dominating the overall sensitivity from several hundreds of GeV up to few TeV. In case of detection a deep
follow up with the Large Sized Telescopes (LST) sensitive down to few tens
of GeV could measure the reprocessed flux and hence the IGMF under the hypothesis that the intrinsic emission does not interfere below 100 GeV. Possible different variability at high (dominated by the intrinsic) and low energies could help to separate the two components.

One of the most interesting features of EHBL is their exceptionally hard
GeV-TeV spectrum, which in most of cases puts the maximum of the SSC component
above few TeV. Here, following \citet{katarzynski05} and \citet{tavecchio09}
we interpret the hard TeV spectrum, together with the extreme X-ray/UV flux
ratio, as the result of the emission from an electron population with a large
minimum Lorentz factor, $\gamma _{\rm min}\simeq 10^5$. As discussed in
\citet{tavecchio09} the exceptionally large $\gamma _{\rm min}$ seem to
challenge existing models for particle acceleration, which usually predicts
power-law non-thermal tails starting from much lower values \citep[e.g.][]{sironi11}. 

Another intriguing point is the apparent stability of the TeV spectrum over
several years (\citealt{aharonian07}, but see also \citealt{aliu14} for the case of 1ES
0229+200 where marginal evidence for mild TeV variability  on a several months/annual time scale is found), which neatly
distinguishes EHBL from HBL that show rather variable TeV emission, down to
timescales of few minutes. As discussed in \citet{tavecchio09} for 1ES
0229+200 the radiative cooling time of electrons at $\gamma _{\rm min}$ is of
the order of 2 years, barely consistent with the observations. However, it is
difficult to avoid other sources of losses, such as adiabatic losses. One
interesting possibility to understand such a stable spectrum \citep[e.g.][]{murase12a,essey11,taylor11} is that the GeV-TeV continuum is not the primary emission
from the source, but instead is reprocessed radiation from cascading processes
spread into the intergalactic space. Electromagnetic cascades could be the
result of the absorption of primary multi-TeV gamma rays or, instead, could be
produced by Bethe-Heitler pair creation or photo-meson reactions involving
ultra-high energy protons accelerated in the source and beamed toward the
Earth. As shown in \citet{tavecchio14}, the physical parameters of the jets are consistent with the requests of the hadronic cascade scenario, both in terms of maximum hadron energy and jet power.
As discussed by \citet{murase12a} and \citet{takami13} an effective
test to distinguish between intrinsic and reprocessed emission is the
observation of photons at several TeV, only possible in the latter case, since
the cascade emission, being produced at lower distance, is less affected by
absorption than the intrinsic one; although exotic processes such as the
photon-axion conversion \citep[e.g.][]{deangelis11} could result in a lower
effective absorption. Due to absorption and reprocessing of primary TeV
  photons a significant contribution to the GeV extragalactig gamma-ray
  background (EGB) could be related to E-HBL
  \citep[see e.g.][]{inoue12,murase12b,ajello14}. We plan to study this issue
  in a forthcoming paper.

Another topic worth of discussion is the position of these EHBLs within the
general BL Lac population and the nature of their parent population. By
construction, the sources we are discussing have very faint radio emission (radio luminosities around $L_r\sim 10^{40}$ erg  s$^{-1}$) indicating very weak radio jets \citep[e.g.][for some
  sources]{giroletti04}. Considering that at least a fraction of the radio flux comes from the beamed jet synchrotron component, we argue that the intrinsic radio luminosity drops below the lower end of the FRI radiogalaxy power range, making  EHBLs suitable candidates for the aligned counterparts of the 
weak radiogalaxies population studied by \citet{baldi09}.

Finally we wish to speculate on the existence of even more extreme EHBL. In
fact it is tempting to suppose the existence of blazars which SED is shifted
toward even higher energies, with the X-ray emission peak  above 10 keV and
high-energy component reaching several tens of TeV. Given the very high energy
of the synchrotron and IC peaks and low flux levels these sources could escape
detection by {\it Swift} and {\it Fermi}. We speculate that possibly the CTA
survey will allow to detect or constrain this hypotetical subclass of the HBL
population in the TeV band. Of course such sources would deeply suffer
  from EBL absorption around their IC emission peak, limiting their detectability at
low redshifts and the detection of the left side of the IC bump below few TeV. Arguably, a relatively strong fraction of their radiated power
would be reprocessed into the GeV band, setting limits to their spatial
density. We will further investigate this issue in a future work.

A generalization of our criterion, for instance relaxing the condition of the known redshift, 
will allow to validate this EHBL selection method on new and larger
samples. This class of sources is especially important for improving our understanding
 about the far-IR EBL, IGMF and EGB. Exploiting at the same time the
 capability of NuStar \citep{harrison13} SKA \citep{carilli04} and CTA
 \citep{acharya13} will be particularly revealing.  Even before the completion
 of CTA, the planned ASTRI/CTA  mini--array \citep{dipierro13,vercellone13}
 could be exploited in this direction.    Actually, whereas the brightest flux
 from this sources is expected to be in  the band up to 1 TeV, where MSTs will
 be the most sensitive instruments of  CTA, still crucial physical information
 is engraved in the high energy tail of the IC bump, in the band best observed with the Small Sized Telescopes.

 For instance, these extreme sources could allow
eventually to compare leptonic and hadronic emission scenarios in a multi-TeV
territory where degeneracy of competing models is significantly
reduced  \citep[see e.g.][]{murase12a}. Opportunity would arise for tests of non standard physics such as violations of the Lorentz invariance \citep{fairbairn14} or
the speculated existence of axion--like particles \citep{meyer14}, once the number of these multi-TeV photon
factories should become greater than now. Therefore EHBL will be
very interesting targets for challenging hot topics in fundamental physics, both with the ASTRI/CTA Mini--Array (improving the
current H.E.S.S. sensitivity above 5 TeV) and with the full array of CTA small
telescopes, which dominates the overall CTA sensitivity above the same
threshold.

\section*{Acknowledgements}

FT and GB acknowledge financial contribution from grant PRIN-INAF-2011.
 This work is based on the publicly available {\it Fermi}/LAT data obtained through the
 Science Support Center (SSC). This research has made use of data obtained
 from the High Energy Astrophysics Science Archive Research Center (HEASARC),
 provided by NASA's Goddard Space Flight Center. We acknowledge the use of
 public data from the Swift data archive. We also acknowledge use of GALEX
 data is made publically available through the Multi--Mission archive at the
 Space Telescope Science Institute (MAST). Part of this work is based on
 archival data, software or on-line services provided by the ASI Science Data
 Center (ASDC).  We thank {\bf the referee Andreas Zech} for comments that helped in
   improving the paper. We also acknowledge constructive comments from
   M. Cerutti, K. Murase, Y. Tanaka and D. Sanchez.

\label{lastpage}
\end{document}